\documentclass[prb,showpacs]{revtex4}
\usepackage{amsmath}
\usepackage{graphicx}
\usepackage{epsfig}
\usepackage{dcolumn}
\usepackage{bm}
\usepackage{dcolumn}
\usepackage{amsfonts}
\usepackage{amssymb}
\providecommand{\U}[1]{\protect\rule{.1in}{.1in}}

\begin{document}

\title{Exact eigenfunctions of FQHE systems \\at fractional filling
factors $\nu=\frac{1}{q}$. I. Formal results.}

\author{Alejandro Cabo$^{*,**}$ and Francisco Claro$^{***}$}

\affiliation{$^{*}$Grupo de F\'isica Te\'orica, Instituto de
Cibern\'etica Matem\'atematica y F\'{\i}sica (ICIMAF),}
\affiliation{Calle E, No. 309, entre 13 y 15, Vedado, La Habana,
Cuba} \affiliation{$^{**}$ Perimeter Institute for Theoretical
Physics, 31 Caroline St., Waterloo, ON, N2L 2Y5, Canada }
\affiliation{$^{\ast\ast\ast}$ Facultad de F\'{\i}sica and Facultad
de Educaci\'{o}n, Pontificia Universidad Cat\'{o}lica de Chile, Av.
Vicu\~{n}a Mackenna 4860, Macul, Santiago, Chile }

\begin{abstract}
Eigenstates of the FQHE hamiltonian problem after to be projected on
the $LLL$ are determined for filling factors $\nu=1/q$, with $q$ an
odd number. The solutions are found for an infinite class of finite
samples in which the Coulomb potential is periodically extended.
Therefore, a thermodynamic limit solution is also identified. The
results suggest the presence of integrability properties in FQHE
systems. The many particle states are simple Slater determinants
constructed with special single particle  states. These orbitals are
defined as powers of order $q$ of $composite\, fermion$ like
wavefunctions associated to a reduced magnetic field $B/q$. At the
same time, those $composite\, fermion$ states were obtained by
factorizing and canceling  $fixed$ position (quasi-momentum
independent) zeros in previously derived exact Hartree-Fock
orbitals. A formula for the energy per particle of the FQHE states
is given for finite samples as well as for the thermodynamic limit
state. As a side result, the same $composite\, fermions$ like
orbitals are employed to construct variational wavefunctions of the
system, showing zeros of order $q$ as two electrons approach each
other, as Laughlin states do. The long range spatial correlation
associated to the starting HF solutions may further reduce the
energy of these states.
\end{abstract}


\maketitle

%
%
%
%
%
%
%

%
%
%
%
%
%
%

\section{Introduction}

The Fractional Quantum Hall effect has been a central theme of investigation
in condensed matter physics for decades
\cite{laughlin,halperin,MacDonald,halrez,kivelson,yoshioka,jain}. Up to now
its understanding rests on Laughlin's variational approach for Landau level
filling fractions of the form $1/q$, for $q$ odd, and its generalization to
general fractional fillings developed by Jain \cite{laughlin, jain}. The wave
functions proposed by these authors are not eigenstates of the system
hamiltonian although their energy is quite close to the true ground state, as
numerical calculations have shown \cite{yoshioka}. An alternative approach to
FQHE originated in the special Hartree-Fock (HF) solutions proposed by F.
Claro \cite{claro1,claro2} many years ago, which although lacking
correlations, capture the essential odd-denominator rule observed in the
fractional quantum Hall regime. Later, further work\ (Refs.
\cite{cabo1,cabo2,cabo3,alamos1,alamos2,ictp}) has shown promising properties
of such solutions, raising the hope that by adding electron correlations one
could approach the true ground state of the system, possibly leading to the
long-standing idea of a weak dynamical breaking of the translation symmetry
\cite{kivelson,halperin,ortolani,chui}.

In this work, in first place, we construct exact eigenstates of the
 restricted to the lowest Landau level ($LLL$) FQHE problem when the
Coulomb potential is periodically extended. These wavefunctions
corresponds to filling factors of the form $\nu=1/q,$ with $q$ an
odd integer. They satisfy periodic boundary conditions, and the
Coulomb potential is defined to have the same periodicity. The
possibility for the determination of these exact wavefunctions is
offered by the use of a special basis of one particle states which
were constructed by factorizing zeros in certain solutions of the HF
problem. Such kind of functions were initially discussed in Ref.
\cite{cabo1,cabo2,cabo3,alamos1,alamos2,ictp}. The starting HF
orbitals showed a structure, in which, from the total number of
zeros (being equal to the number of quantum fluxes piercing the
periodicity region) only a fraction $q$ had positions depending of
the quasimomenta quantum numbers of the orbitals. The rest of
$(q-1)/q$ zeros were situated at the origin of coordinates. Those
$fixed$ zeros where able to be factorized and canceled, by then
leaving as a result a set of $composite$ $fermion$ like orbitals,
associated to a reduced magnetic field $B/q$. This property directly
led to the idea of the possible physical relevance of special single
particle orbitals constructed as powers of order $q$ of such states
(to be called below: $q$ $product$ states). It can be noted that
those products regain the number of zeros required by the actual
magnetic field value. Further motivation of the line of search under
consideration came after checking that the set of orbitals
constructed for all the quasimomenta values, became closed under the
action of the magnetic translations defining the periodically
extended Coulomb potential. \ The extended potential is defined to
have periods coinciding with the same assumed here von Karman
periodicity lengths of the system: $L_{1}/q$ and $L_{2}/q$. \ The
mentioned closure property, then allows to directly show that the
determinant functions formed with the $q$ $product$ states are in
fact exact solutions of the Schrodinger equation. A simple formula
for the exact eigen-energy of the system also follows. Afterwards,
since the solution was determined for arbitrary but finite size of
the systems, the thermodynamic limit of problem can also be
determined.

\ The decision about whether these exact eigenstates of the FQHE
systems are excited or the ground state ones, will be further
considered in a work now in progress (paper II). That is, we do not
argue that the presented exact solutions represent the ground state
of the FQHE\ system, which the variational Jain-Laughlin approaches
and the exact diagonalization results very well approximate.
Occasionally, the determined states could result to be excited ones,
showing even higher energy per particle than the Laughlin
wavefunctions. However, even in that case, the wavefunctions could
describe alternative phases of the 2DEG which having physical
interest. This can be the situation, because the experiments in FQHE
systems are performed over a wide range of conditions. By example, \
particularly important kind of measurements are related with finite
temperature experiences \cite{pam1,pam2,pam3,pam4}. In them, the
existence of phase transitions between states of the 2DEG when the
temperature changes has been detected. Such transitions are related
with the relaxation of the FQHE systems to states which can not be
described by just few excitation over the ground state, but should
correspond two radically different ground states, showing higher
energy per particle values. This circumstance indicates the physical
interest of the investigated exact eigenstates after taken in the
thermodynamic limit. On another hand, if the exact solutions could
occasionally correspond to the ground states, their physical
interest will be clear.

It should be noticed that this initial part of the work contains a
large amount of formal and technical developments which are however,
required for arguing the exact nature of the wavefunctions.
Therefore, in order to avoid a disproportionate larger paper, we
plan to publish the results as a series of two articles: papers I
and II. The present one (to be referred as paper I) will contain the
formal derivation of the exact eigenstates. In the second work of
the series (paper II) we plan to present the evaluations of the
energy per particle and pair correlation functions of the here
discussed eigenstates. Such results should define whether those
wavefunctions correspond to phases that only can be realized at high
temperatures, or on another hand, can also describe the exact ground
states of the FQHE problem.

It can be remarked that the connection of the exact states discussed
here with the solutions of the exactly solvable model defined in a
thin torus derived in Refs.\cite{hans1,hans2,hans3,hans4,hans5} is
also planned to be examined in the next paper II.

In connection with the structure of the zeros of the obtained
solutions, it can be underlined that the almost exact states
determined in Refs. \cite{peeters1,peeters2, peeters3} for Quantum
Dots in FQHE configurations, indicate that the zeros of the exact
wavefunctions are not of order $q$ as it occurs in the Laughlin and
Jain variational states. The determinant nature of the functions
studied here, seems to be compatible with this property. However,
this point is expected to be more clarified also in the planned
paper II.

Another  objective of this work is treated in an  Appendix. There,
the same special one particle functions are also employed to
construct ansatz for low energy eigenfunctions. They are defined as
powers of order $q$ of determinant like states associated to samples
having a magnetic field reduced to a fraction $q$ of the real
external one. The powers $q$ entering in the construction, ensure
that the proposed states show the required number of zeros
associated to the external field in the periodicity region. This
form of definition directly implies that the zeros of the
wavefunction at nearly
coinciding coordinates $z_{i}$ and $z_{j}$ are of the form $(z_{i}-z_{j})^{q}%
$. Thus, the advanced states  show a similar short range behavior as
the Laughlin states which thanks to such a property have the status
of being the best variational state proposals up to now. However,
since the starting point for the definition of the here advanced
ansatz, are the HF one particle states, the expectation emerges,
about that the long range correlations incorporated in the HF
solutions, will remains to be effective in the new wavefunctions, by
occasionally helping to lower the energies slightly more.

The paper will proceed as follows. In Section II the wavefunction
corresponding to an exact eigenvalue of the periodically extended
Coulomb Hamiltonian reduced to the $LLL$, will be presented.  Also,
a formula for the energy per particle of the FQHE state is derived
and  the thermodynamic limit of the state is analyzed. Section III
exposes the conclusions and a summary of the results. In Appendices
A and B the definitions and auxiliary technical derivations are
given for the sake of completeness and bookkeeping purposes. The
construction of the ansatz wavefunctions for the ground state of
FQHE systems at filling factors $1/q$ is discussed in Appendix C.

\section{Exact solutions of the FQHE problem }

Let us consider a 2-dimensional electron gas (2DEG)in the presence
of a perpendicular magnetic field $\mathbf{B}$ such that a fraction
$\nu=1/q$ \ of the lowest Landau level $(LLL)$ is filled, where $q$
is an odd integer. \ The electronic systems will be periodically
extended in what follows to satisfy magnetic periodic von Karman
conditions \ in the \ lattice generated by the vectors \cite{halrez}
\[
\frac{\mathbf{L}_{1}}{q}=\frac{N\ \mathbf{a}_{1}}{q},\text{ \ \ }%
\frac{\mathbf{L}_{2}}{q}=\frac{N\ \mathbf{a}_{2}}{q}.
\]

Here $N$ \ is an even  multiple of $q,$ such that precisely one flux
quanta pierces the cell of sides $\mathbf{L}_{1}/(qN)$ and
$\mathbf{L}_{2}/N$ \ \ The unit cell vectors $\mathbf{a}_{1}$ and
$\mathbf{a}_{2}$ define the smaller spatial magnetic translations
under which the HF orbitals motivating
this work, were eigenfuntions \cite{cabo1,cabo2,cabo3,alamos1,alamos2,ictp}%
\ (see Appendix A for further definitions and conventions).

As it was mentioned in the Introduction, the Coulomb potential between
particles will be periodically extended to be invariant under translations in
the same lattice defining the von Karman periodicity of the electron
wavefunctions \
\begin{align}
\mathbf{R}_{q}  &  =R_{q,1}\frac{\text{ }N\mathbf{a}_{1}}{q}+R_{q,2}\text{
}\frac{N\mathbf{a}_{2}}{q}\nonumber\\
&  =R_{q,1}\frac{\text{ }\mathbf{L}_{1}}{q}+R_{q,2}\text{ }\frac
{\mathbf{L}_{2}}{q}, \label{Rq}%
\end{align}
in which $R_{q,1}$ and $R_{q,2}$ are arbitrary integers. Then, the
extended Coulomb potential has the expression
\begin{equation}
V_{C}(\mathbf{x}-\mathbf{x}^{\prime})=\sum_{\ \mathbf{R}_{q}}\frac{e^{2}%
}{|\mathbf{x}-\mathbf{x}^{\prime}+\mathbf{R}_{q}|},
\end{equation}
which is periodic in the lattice $R_{q}$. After projection on the
$LLL$, the Coulomb Hamiltonian can be written in the Landau gauge,
in the form (See Appendix A and Ref. [\onlinecite{yoshioka}])
\begin{align}
H  &  =\frac{1}{A_{c}}\sum_{\mathbf{l}}\sum_{i<j}\frac{2\pi e^{2}r_{o}^{2}%
}{\ |\mathbf{l}|}\exp(-\frac{l\text{ }l^{\ast}}{2r_{o}^{2}})\text{
\ }T(\ \mathbf{l},z_{i}^{\ast})\text{ }T(-\ \mathbf{l},z_{j}^{\ast
}),\label{hamq}\\
A_{c}  &  =\frac{\mathbf{n}\cdot\mathbf{L}_{1}\times\mathbf{L}_{2}}{q^{2}},
\end{align}
where the magnetic translation operators in an arbitrary vector $\mathbf{v}%
$\ are given by (See Appendix A)
\begin{equation}
T(\mathbf{v},z^{\ast})=\text{ }\exp(-\frac{(\mathbf{v})_{2}}{r_{o}^{2}}%
(\frac{v^{\ast}}{2}+i\text{ }z^{\ast}))\exp(v^{\ast}\frac{\partial}{\partial
z^{\ast}}),
\end{equation}
and the vectors $\mathbf{l}\ \ $are defined in terms of the normal vector
$\mathbf{n}$ and the discrete reciprocal lattice momenta (associated to the
von Karman lattice) $\mathbf{k}$ as
\begin{align}
\mathbf{l}  &  =r_{o}^{2}\mathbf{n\times k,}\label{lvalues}\\
\mathbf{k}  &  =\frac{q_{1}}{\frac{N}{q}}\,\,\mathbf{s}_{1}+\frac{q_{2}}%
{\frac{N}{q}}\mathbf{s}_{2},\\
q_{1},q_{2}  &  =-\infty,..,-2,-1,0,1,2,...,\infty,
\end{align}
in which the vectors $\mathbf{s}_{1}$ and $\mathbf{s}_{2},$ are
defined in Appendix A. Technical definitions and a derivation of the
expression of the above periodically extended Hamiltonian can be
also found in Appendix A.

Let us now consider the following Slater wavefunction for a system of
$N_{e}=(\frac{N}{q})^{2}$ electrons,
\begin{align}
\Psi(z_{1}^{\ast},z_{2}^{\ast},...z_{N_{e}})  &  =Det[\Theta_{\mathbf{l}_{i}%
}(z_{j}^{\ast})], \label{fi}\\
Det[\Theta_{\mathbf{l}_{i}}(z_{j}^{\ast})]  &  =\sum_{P}(-1)^{p}%
\Theta_{\mathbf{l}_{1^{P}}}(z_{1}^{\ast})\Theta_{\mathbf{l}_{2^{P}}}%
(z_{2}^{\ast})...\Theta_{\mathbf{l}_{N_{e}^{P}}}(z_{N_{e}}^{\ast}),
\end{align}
where $p$ is the order of a given permutations $P$ of the set of numbers
$\{{1,2,3,...,N_{e}}\},$ defined by $P\{{1,2,3,...,N_{e}\}}=\{{1^{P}%
,2^{P},3^{P},...,N_{e}^{P}\}}$. The single particle orbitals
employed to construct the determinant have the expressions
\begin{align}
\Theta_{\mathbf{k}}(z^{\ast})  &  =\exp(-i\text{ }\varkappa\text{ }z^{\ast
}){\Large (}\chi_{_{\frac{\mathbf{k}}{q}}}(z^{\ast}){\Large )}^{q}, \label{11}\\
\mathbf{l}  &  =r_{o}^{2}\mathbf{n}\times\mathbf{k}=-(\frac{q_{2}}%
{N})\mathbf{a}_{1}+(\frac{q_{1}}{N})\mathbf{a}_{2},\\
q_{1}  &  =-\frac{N}{2},-\frac{N}{2}+1,...,0,...,\frac{N}{2}-1,\\
q_{2}  &  =-\frac{N}{2},-\frac{N}{2}+1,...,0,...,\frac{N}{2}-1,
\end{align}
and $\varkappa=-\pi q(q-1)/a$. \ Note that the $quantum$ $numbers$
$\mathbf{l}$ $\ $\textbf{(}or their equivalent $\mathbf{k}$\textbf{)} of the
set of one particle functions $\Theta_{\mathbf{l}}$ take $N_{e}=N^{2}/q^{2}$
values. That is, the subspace formed by this set of functions is filled in the
proposed may particle state. Since the total number of flux quanta traversing
the\ unit cell of the periodicity region is $\Phi=q\ \frac{N^{2}}{q^{2}},$ the
filling fraction of the ansatz state is just $\nu=1/q.$ \ The functions $\chi$
entering the above definition are defined as follows,
\begin{align}
\chi_{\mathbf{q}}(z^{\ast})  &  =\exp(i\frac{\mathbf{q}.\mathbf{a}_{1}}%
{a}\ z^{\ast})\times\prod_{R}\vartheta_{1}(\frac{\pi}{L}(z^{\ast}-R^{\ast
}-C_{\mathbf{q}})|-\tau^{\ast}),\\
C_{\mathbf{q}}  &  =\frac{a}{2\pi}(\mathbf{q}.\mathbf{a}_{1}\tau^{\ast
}-\mathbf{q}.\mathbf{a}_{2})+\frac{q\text{ }\tau^{\ast}a}{2}, \label{composf}%
\end{align}
where the vectors $R$ (note that the vectors will be designed either by their
boldface symbols of by usual symbols associated to their complex
representations: See Appendix A) defining the product, and $\mathbf{q}$\ \ are
given by
\begin{align}
R=r_{1}a+r_{2}a_{2},\,R^{\ast}  &  =r_{1}a+r_{2}a_{2}^{\ast},\,a_{2}%
=a\exp(i\frac{\pi}{3}),\\
r_{1},r_{2}  &  =-\frac{N}{2},...,0,...,\frac{N}{2}-1,\\
\mathbf{q}  &  =\frac{q_{1}}{N}\mathbf{s}_{1}+\frac{q_{2}}{N}\mathbf{s}_{2},\\
q_{1},q_{2}  &  =-\frac{N}{2},-\frac{N}{2}+1,...,0,...,\frac{N}{2}-1,\\
\tau^{\ast}  &  =\frac{a_{2}^{\ast}}{a},\nonumber
\end{align}
and $\vartheta_{1}$ is the elliptic Theta function
\begin{align}
\vartheta_{1}(u|\alpha)  &  =2\sum_{n=1}^{\infty}(-1)^{n}\exp(i\pi
\alpha(n-\frac{1}{2})^{2})\sin((2n-1)u),\\
\operatorname{Im}(\alpha)  &  >0.
\end{align}
\ \ The above functions $\chi$ are representations in terms of the
Elliptic Theta functions of the $composite$ $fermion$ like states
derived from the exact HF single particle solutions discussed in the
Ref \ \cite{cabo1,cabo2,cabo3,alamos1,alamos2,ictp}, after
factorizing and canceling zeros situated at momentum independent
positions. Note that the momenta argument of the functions $\chi$ in
he definition of the orbitals in (\ref{11}) has been divided by $q$.
 this selection was heuristically motivated by considering the power $q$
of the $composite\, fermion$ should be expected to determine a
momentum of the single particle wavefunction being $q$ times greater
than the one associated to each of the  $q$ $composite\, fermion$
constituent blocks.

 It should be noted, that the case of \ HF states associated to a crystal
periodicity showing one electron per unit cell $(\gamma=1)$ and
$\nu=1/q$ was considered here. In this sense, it can be underlined
that a similar discussion to the one given here seems possible to be
done for the case $(\gamma=\frac{1}{2})$ and $\nu=1/q.$

In the next subsection we will show that the functions
$\Theta_{\mathbf{l}}$ satisfy the following closure properties under
a magnetic translation in an arbitrary vector $\ $of the form
$\mathbf{l'}$:
\begin{equation}
T(\ \mathbf{l}^{\prime},z^{\ast})\text{ }\Theta_{\mathbf{l}}(z^{\ast}%
)=\exp(\Phi_{\mathbf{l}}(\mathbf{l}^{\prime}))\text{ }\Theta_{\lbrack
\mathbf{l}-\mathbf{l}^{\prime}]_{red}}(z^{\ast}), \label{fase1}%
\end{equation}
where $[\mathbf{l}-\mathbf{l}^{\prime}]_{red}$ \ means the vector equivalent
to $\mathbf{l}-\mathbf{l}^{\prime}$ modulo an element of the lattice generated
by the unit cell vectors $\mathbf{a}_{1}$ and $\ \mathbf{a}_{2}$. The
satisfaction of the periodic boundary conditions in the von Karman periodicity
region of the functions $\Theta_{\mathbf{l}}$ will be shown in Appendix B.

\subsection{Action of the magnetic translations on the functions
$\Theta_{\mathbf{l}}(z^{\ast})$}

Let us now argue that the space generated by the set of functions
$\Theta_{\mathbf{l}}(z^{\ast})$ for all $\mathbf{l}$\textbf{\ }values of the
quantum numbers is closed under the action of arbitrary magnetic translations
in any vector of the form $\ \mathbf{l}^{\prime}$ which are the spacial shift
defining the Hamiltonian in (\ref{hamq}). \ This action can be expressed as follows%

\begin{equation}
T(\mathbf{l}^{\prime},z^{\ast})\text{ }\Theta_{\mathbf{l}}(z^{\ast}%
)=\exp(-\frac{i(\mathbf{l}^{\prime}\mathbf{)}_{2}\text{ }}{r_{o}^{2}}%
(\frac{l^{\prime\ast}}{2}+z^{\ast})-i\varkappa(z+l^{\prime\ast}))\times
{\large (}\chi_{\frac{_{\mathbf{k}}}{q}}(z^{\ast}+l^{\prime\ast}%
){\large )}^{q}. \label{transf}%
\end{equation}

But, the appearing power $q$ of the functions $\chi$ can be explicitly written
in the form%

\begin{equation}
{\large (}\chi_{\frac{_{\mathbf{k}}}{q}}(z^{\ast}+l^{\prime\ast}%
){\large )}^{q}=\exp(\frac{i(\mathbf{l)}_{2}\text{ }}{r_{o}^{2}}l^{\prime\ast
}+\frac{i\text{ }(\mathbf{l)}_{2}\text{ }}{r_{o}^{2}}z^{\ast})\left(
\prod_{R}\theta_{1}(\frac{\pi}{L}(z^{\ast}-R^{\ast}-\frac{q}{2}a_{2}^{\ast
}-(l^{\ast}-l^{\ast\prime}))|-\tau\ast)\right)  ^{q}.
\end{equation}

After substituting this expression in (\ref{transf}), it follows
\begin{align}
T(\ \mathbf{l}^{\prime},z^{\ast})\text{ }\Theta_{\mathbf{l}}(z^{\ast})  &
=\exp(-\frac{i(\mathbf{l}^{\prime}\mathbf{)}_{2}\text{ }}{r_{o}^{2}}%
\frac{l^{\ast\prime}}{2}+\frac{i(\mathbf{l)}_{2}\text{ }}{r_{o}^{2}}%
l^{\prime\ast}-i\varkappa l^{\prime\ast})\times\nonumber\\
&  \exp(-i\varkappa\text{ }z^{\ast})\left(  \exp(\frac{i\text{ }%
(\mathbf{l-l}^{\prime}\mathbf{)}_{2}\text{ }}{qr_{o}^{2}}z^{\ast})\prod
_{R}\theta_{1}(\frac{\pi}{L}(z^{\ast}-R^{\ast}-\frac{q}{2}a_{2}^{\ast
}-q(l^{\ast}-l^{\ast\prime}))|-\tau\ast)\right)  ^{q},\label{act1}\\
&  =\exp(-\frac{i(\mathbf{l}^{\prime}\mathbf{)}_{2}\text{ }}{r_{o}^{2}}%
\frac{l^{\ast\prime}}{2}+\frac{i(\mathbf{l)}_{2}\text{ }}{r_{o}^{2}}%
l^{\prime\ast}-i\varkappa l^{\prime\ast})\times\Theta_{\mathbf{l-l}^{\prime}%
}(z^{\ast}).
\end{align}

As it can be noted, the last line in the above equation is the same analytic
expression defining the function $\Theta_{\mathbf{l}}$, but in which the
shifted quantum number value $\mathbf{l-l}^{\prime}$, may lay outside the set
defining the allowed values of such quantum numbers. However, the difference
$\mathbf{l-l}^{\prime}$ can be expressed in terms of a proper value of a
quantum number modulo a linear combination of some basic quasi periods of the
functions $\Theta.$ For seeing this property, let us note that $l^{\ast}%
-l^{\prime\ast}$ can always be written as
\begin{equation}
l^{\ast}-l^{\prime\ast}=[l^{\ast}-l^{\prime\ast}]_{red}-t_{1}a-t_{2}%
a_{2}^{\ast},
\end{equation}
where $[l^{\ast}-l^{\prime\ast}]_{red}$ \ has the quantum number
form
\begin{align}
\lbrack l^{\ast}-l^{\prime\ast}]_{red}  &  =\frac{r_{1}}{\frac{N}{q}}%
a+\frac{r_{2}}{\frac{N}{q}}a_{2}^{\ast},\\
r_{1},r_{2}  &  =-\frac{N}{2q},-\frac{N}{2q}+1,....,0,...\frac{N}%
{2q}-1,\nonumber
\end{align}
and $t_{1}$ , $t_{2}$ are defined as functions of $\ l^{\ast}$ and
$\ l^{\prime \ast}$ as the unique pair of integer numbers required
to express $[l^{\ast}-l^{\prime\ast}]_{red}$ in the above form. The
definitions of $\ \mathbf{l}$ and $\mathbf{l}^{\prime}$ permits to write for them%
\begin{align}
t_{1}  &  =-Floor[\frac{r_{1}-r_{1}^{\prime}+\frac{N}{2q}}{\frac{N}{q}}%
]\frac{N}{q},\nonumber\\
t_{2}  &  =-Floor[\frac{r_{2}-r_{2}^{\prime}+\frac{N}{2q}}{\frac{N}{q}}%
]\frac{N}{q}, \label{t1t2}%
\end{align}
where the operation $Floor[n]$ as usual, is defined as the nearest integer
number being lower or equal that the real number $n.$ Note that the indices
$r_{1}^{\prime},r_{2}^{\prime}$ take arbitrarily large values because they
define the translations entering in the periodic expansion of the Coulomb
potential. On another hand the values $r_{1},r_{2}$ only define the set
quantum numbers.

Then, we can write for the main quantity entering in the arguments of the
Theta functions as follows \
\begin{equation}
(l^{\ast}-l^{\prime\ast})+\frac{q}{2}a_{2}^{\ast}+R^{\ast}=[l^{\ast}%
-l^{\ast\prime}]_{red}+\frac{q}{2}a_{2}^{\ast}+R^{\ast}-t_{1}a-t_{2}%
a_{2}^{\ast}.
\end{equation}
Therefore, the shifts in the vectors $\ l^{\prime\ast}$ produce a
new argument which coincides with an argument of another function in
the considered set, plus a shift in an integer number of \ the unit
cell vectors $\mathbf{a}_{1}$ and $\mathbf{a}_{2}$. \ However, the
functions $\Theta_{\mathbf{l}}$ under consideration satisfy exact
recurrence relations under such displacements. \ They are direct
consequences of the usual transformation properties of the Theta
functions under shifts in their quasiperiods. The derivation of
these relations is presented in Appendix B. \ In order to start
making use of this property, the expression (\ref{act1}) for the
action of the translations on the functions \ $\Theta_{\mathbf{l}}$
can be written in the
following form%

\begin{align}
T(\mathbf{l}^{\prime},z^{\ast})\text{ }\Theta_{\mathbf{l}}(z^{\ast})  &
=\exp(-\frac{i(\mathbf{l}^{\prime}\mathbf{)}_{2}\text{ }}{r_{o}^{2}}%
\frac{l^{\ast\prime}}{2}+\frac{i(\mathbf{l)}_{2}\text{ }}{r_{o}^{2}}%
l^{\prime\ast}-i\varkappa l^{\prime\ast})\times\\
&  \exp(-i\varkappa\text{ }z^{\ast})\exp(\frac{i\text{ }(\mathbf{l-l}^{\prime
}\mathbf{)}_{2}\text{ }}{r_{o}^{2}}z^{\ast})\times\left(  \Omega(v^{\ast
}(l^{\prime},l,z^{\ast})+t_{1}a+t_{2}a_{2}^{\ast})\right)  ^{q},\nonumber\\
\Omega(w^{\ast})  &  =\prod_{R}\theta_{1}(\frac{\pi}{L}(w^{\ast}-R^{\ast
})|-\tau^{\ast}),\\
v^{\ast}(l^{\prime},l,z^{\ast})  &  =z^{\ast}-\frac{q}{2}a_{2}^{\ast}%
-[l^{\ast}-l^{\ast\prime}]_{red}.
\end{align}
Then, we can make use of the relation (\ref{aux1}) in Appendix B, for the
above defined functions $\Omega$%

\begin{equation}
\Omega(v^{\ast}+m^{\ast})=\exp(i\pi m_{2}+2i\,\pi\,m_{2}\frac{v}{a}^{\ast
}+i\,\pi\,\tau^{\ast}m_{2}(m_{2}+1)\,)\,\Omega(v^{\ast}),
\end{equation}
in which the vector $\mathbf{m}$ for our situation should be chosen as
$\ \mathbf{m}=m_{1}\mathbf{a}_{1}+m_{2}\mathbf{a}_{2}$ being equal to
$t_{1}\mathbf{a}_{1}+t_{2}\mathbf{a}_{2}$. Applying this formula for
transforming the product $\mathcal{P}_{1}$ of \ $\Omega$ \ functions appearing
in the previous expression:%

\begin{align}
\mathcal{P}_{1}  &  =\left(  \Omega(v^{\ast}(l^{\prime},l,z^{\ast}%
)+t_{1}a+t_{2}a_{2}^{\ast})\right)  ^{q}\\
&  =\left(  \prod_{R}\theta_{1}(\frac{\pi}{L}(v^{\ast}(l^{\prime},l,z^{\ast
})+t_{1}a+t_{2}a_{2}^{\ast}-R^{\ast})|-\tau^{\ast})\right)  ^{q},\nonumber
\end{align}
we can write the shifted $\Omega$ functions as multiplicative factors of the
original ones as follows
\begin{equation}
\Omega(v^{\ast}(l^{\prime},l,z^{\ast})+t_{1}a+t_{2}a_{2}^{\ast})=\exp(i\pi
t_{2}+2i\pi t_{2}\frac{v^{\ast}(l^{\prime},l,z^{\ast})}{a}+i\pi t_{2}%
(1+t_{2}))\Omega(v^{\ast}(l^{\prime},l,z^{\ast})).
\end{equation}
Then, $\mathcal{P}_{1}$ \ have the expression%

\begin{align}
\mathcal{P}_{1}  &  =\exp(i\pi qt_{2}+2i\pi qt_{2}\frac{v^{\ast}(l^{\prime
},l,z^{\ast})}{a}+i\pi qt_{2}(1+t_{2}))\times\\
&  (\left(  \Omega(v^{\ast}(l^{\prime},l,z^{\ast}))\right)  ^{q}.\nonumber
\end{align}

Therefore, the action of the translations on the function $\Theta_{\mathbf{l}%
}$ \ can be expressed as%
\begin{align}
T(\mathbf{l}^{\prime},z^{\ast})\text{ }\Theta_{\mathbf{l}}(z^{\ast})  &
=\exp(-i\frac{(\mathbf{l})_{2}^{\prime}\text{ }l^{\prime\ast}}{2r_{o}^{2}%
}+i\frac{(\mathbf{l})_{2}\text{ }l^{\prime\ast}}{r_{o}^{2}}-i\varkappa
l^{\prime\ast}-i\varkappa z^{\ast})\nonumber\\
&  \exp\left(  \frac{i}{r_{o}^{2}}([\mathbf{l}-\mathbf{l}^{\prime}]_{red}%
)_{2}z^{\ast}-\frac{i}{r_{o}^{2}}(t_{1}\frac{\mathbf{a}_{1}}{q}+t_{2}%
\frac{\mathbf{a}_{2}}{q})_{2}\text{ }z^{\ast}\right) \nonumber\\
&  \exp(i\pi q\,t_{2}+\frac{2\pi iqt_{2}}{a}(z^{\ast}-\frac{qa\tau^{\ast}}%
{2}-[l\mathbf{-}l^{\prime}]_{red})+i\pi\tau^{\ast}qt_{2}(1+t_{2}%
))\times\nonumber\\
&  \left(  \prod_{R}\theta_{1}(\frac{\pi}{L}(z^{\ast}-R^{\ast}-\frac{q}%
{2}a_{2}^{\ast}-[l^{\ast}-l^{\ast\prime}]_{red})|-\tau\ast)\right)  ^{q}.
\end{align}
After employing the following identity
\[
(\mathbf{a}_{2})_{2}a=q\times2\pi r_{o}^{2},
\]
which reflects that exactly $q$ flux quanta pass through the unit
cell formed by the vectors $\mathbf{a}_{1}$ and $\mathbf{a}_{2}$, by
also noticing that
$(t\frac{\mathbf{a}_{1}}{q})_{2}=0$, because the vector $\ \mathbf{a}%
_{1}=a(1,0,0)$, the following closure relation arises \ \ \ %

\begin{align}
T(\mathbf{l}^{\prime},z^{\ast})\text{ }\Theta_{\mathbf{l}}(z^{\ast})  &
=\exp(-i\frac{(\mathbf{l}^{\prime})_{2}\text{ }l^{\prime\ast}}{2r_{o}^{2}%
}+i\frac{(\mathbf{l})_{2}\text{ }l^{\prime\ast}}{r_{o}^{2}}-i\varkappa
l^{\prime\ast})\times\nonumber\\
&  \exp(i\pi q\,t_{2}-\frac{2\pi iqt_{2}}{a}(\frac{qa\tau^{\ast}}%
{2}+[l\mathbf{-}l^{\prime}]_{red})+i\pi\tau^{\ast}qt_{2}(1+t_{2}%
))\times\nonumber\\
\exp(-i\varkappa &  z^{\ast})\exp(\frac{i}{r_{o}^{2}}([\mathbf{l-l}^{\prime
}]_{red})_{2}z^{\ast})\left(  \prod_{R}\theta_{1}(\frac{\pi}{L}(z^{\ast
}-R^{\ast}-\frac{q}{2}a_{2}^{\ast}-q[l^{\ast}-l^{\ast\prime}]_{red})|-\tau
\ast)\right)  ^{q}\nonumber\\
&  =\exp(\Phi_{\mathbf{l}}(\mathbf{l}^{\prime}))\ \Theta_{\lbrack
\mathbf{l-l}^{\prime}]_{red}}(z^{\ast}). \label{relat}%
\end{align}

The space independent multiplicative factor has the formula%

\begin{align}
\exp(\Phi_{\mathbf{l}}(\mathbf{l}^{\prime}))  &  =\exp(-i\frac{(\mathbf{l}%
^{\prime})_{2}\text{ }l^{\prime\ast}}{2r_{o}^{2}}+i\frac{(\mathbf{l}%
)_{2}\text{ }l^{\prime\ast}}{r_{o}^{2}}-i\varkappa l^{\prime\ast}%
)\times\nonumber\\
&  \exp(i\pi q\,t_{2}-\frac{2\pi iqt_{2}}{a}(\frac{qa\tau^{\ast}}%
{2}+[l\mathbf{-}l^{\prime}]_{red})+i\pi\tau^{\ast}qt_{2}(1+t_{2})).
\end{align}

The integer numbers $t_{1}$ and $t_{2}$ were fixed before in (\ref{t1t2}) as
functions of $\ \mathbf{l,}$ $\mathbf{l}^{\prime}$ by the condition of
enforcing the vector $[\mathbf{l-l}^{\prime}]_{red}$ to be an allowed quantum
number. \ \

\subsection{The eigenstates of the FQHE in the $LLL$ for projected Coulomb interaction}

Let us consider now the central point in this work. Note that the
action of an arbitrary translation entering in the definition \
(\ref{hamq}) of the periodic potential, on any of the functions
$\Theta_{\mathbf{l}}$, reduces to a permutation of the set of
quantum numbers times a space independent exponential factor. Thus,
the defined subspace of functions on the $LLL$ remains invariant
under the class of translations appearing in the expansion of the
interaction potential.

Let us recall here that the number of inequivalent translations
\ in vectors $l^{\prime}%
$\textbf{\ }is $N^{2}/q^{2}$, which is $q$ \ times smaller than the
number of flux quanta $q(N^{2}/q^{2})$ passing though one von Karman
periodicity cell of area $\mathbf{n.L}_{1}\mathbf{\times
L}_{2}/q^{2}.$  By inequivalent translations we mean the whole set
of translations obtained for arbitrary values of the   $quantum \,
\, number$ $ \mathbf{l}$. Note that due to the closure relation, the
translations differing from tho e inequivalent ones in an arbitrary
linear combination with a integral coefficients of the vectors
$\textbf{a}_1$ and $\textbf{a}_2$, reduce to inequivalent
 translation times a multiplicative factor.\ For $q=1$ the set of translations of the form
$l^{\prime}$ transforms the $LLL$ level in itself since the number
of translations coincides with the number of independent function in
this level and any translation implements a permutation of the
functions. However, for $q>1$, \ the number of translations
$l^{\prime}$ \ is\ again equal to the number of functions
$\Theta_{\mathbf{l}},$ but this common number is $q$ times smaller
than number of independent functions in the $LLL$ (equal to the
number of flux quanta passing the region of area
$\mathbf{n.L}_{1}\mathbf{\times L}_{2}/q^{2}).$ \

\ Let us now \ inspect how the Hamiltonian of the system acts over
the wavefunction $\Psi$ defined in (\ref{fi}). Through the help of
relation \ (\ref{fase1}) the result of this action can be written in
the form
\begin{align}
H\text{ }\Psi(z_{1}^{\ast},z_{2}^{\ast},...,z_{N_{e}}^{\ast}) &  =\frac
{1}{A_{c}}\sum_{i<j}\sum_{\mathbf{l}}\frac{2\pi e^{2}r_{o}^{2}}{|\mathbf{l}%
|}\exp(-\frac{l\text{ }l^{\ast}}{2r_{o}^{2}})\text{ \ }T(\mathbf{l}%
,z_{i}^{\ast})\text{ }T(-\mathbf{l},z_{j}^{\ast})\times\nonumber\\
&  \sum_{P}(-1)^{p}\Theta_{\mathbf{l}_{1^{P}}}(z_{1}^{\ast})\Theta
_{\mathbf{l}_{2^{P}}}(z_{2}^{\ast})...\Theta_{\mathbf{l}_{N_{e}^{P}}}%
(z_{N_{e}}^{\ast}),\nonumber\\
&  =\frac{1}{A_{c}}\sum_{i<j}\sum_{\mathbf{l}}\frac{2\pi e^{2}r_{o}^{2}%
}{|\mathbf{l}|}\exp(-\frac{l\text{ }l^{\ast}}{2r_{o}^{2}})\times\nonumber\\
&  \sum_{P}(-1)^{p}\Theta_{\mathbf{l}_{1^{P}}}(z_{1}^{\ast})\Theta
_{\mathbf{l}_{2^{P}}}(z_{2}^{\ast})...\exp(\Phi_{\mathbf{l}_{i^{P}}%
}(\mathbf{l}))\Theta_{\mathbf{l}_{i^{P}}-\mathbf{l}}(z_{i}^{\ast
})...\nonumber\\
&  \text{ \ \ \ \ \ \ \ \ \ \ \ \ \ \ }\exp(\Phi_{\mathbf{l}_{j^{P}}%
}(-\mathbf{l}))\Theta_{\mathbf{l}_{j^{P}}+\mathbf{l}}(z_{j}^{\ast}%
)...\Theta_{\mathbf{l}_{N_{e}}}(z_{N_{e}}^{\ast}).\label{hphi}%
\end{align}
Let us now recall that the action of the Hamiltonian on an antisymmetric
function of the coordinates gives again an antisymmetric function of the
coordinates. Therefore, it should be invariant under the operation of adding
its expression for all permutations of the arguments $z_{1}^{\ast},z_{2}%
^{\ast},...z_{N_{e}}^{\ast}$ , after multiplication by the order of
the permutations, and a division by $N_{e}!.$ However, it can be
noted that the action of the two translations involved produce other
functions of the same set, showing modified momenta quantum numbers
and being multiplied by  constant exponential factors. Therefore,
when the increment$\mathbf{\ l}$ is unable to transform both quantum
numbers $\mathbf{l}_{i^{P}}$ and $\mathbf{l}_{j^{P}}$ between
themselves modulo a vector of the form $m_{1}\mathbf{a}_{1}+m_{2}$
$\mathbf{a}_{2}$ ($m_{1}$ and $m_{2}$ integers), the corresponding \
antisymmetrized term in the sum over permutations in equation
(\ref{hphi}) will vanish. This is a result of the fact that, in the
mentioned situation, there will be two coincident states in the
Slater determinant resulting from the above described auxiliary
symmetrization process. \ Therefore, these properties imply the
following eigenvalue equation to hold
\begin{align}
H\text{ }\,\Psi &  =E_{p}\Psi\,,\label{eigen}\\
E_{p} &  =-\frac{1}{A_{c}}\sum_{i<j}\sum_{\mathbf{l}}\sum_{P}\frac{1}{N_{e}%
!}\frac{2\pi e^{2}r_{o}^{2}}{|\mathbf{l}|}\exp(-\frac{l\text{ }l^{\ast}%
}{2r_{o}^{2}})\times\nonumber\\
&  \exp(\Phi_{\mathbf{l}_{i^{P}}}^{(1)}(\mathbf{l})+\Phi_{\mathbf{l}_{j^{P}}%
}^{(1)}(-\mathbf{l}))\times\text{ }\delta^{(K,P)}(\mathbf{l}-(\mathbf{l}%
_{i^{P}}-\mathbf{l}_{j^{P}})),
\end{align}
where $\ \ \delta^{(K,P)}(\mathbf{l})=1$ if \ $\mathbf{l}=0$\ modulo a lattice
vector of the form $m_{1}\mathbf{a}_{1}+m_{2}\mathbf{a}_{2}$, and vanishes
otherwise. Equation (\ref{eigen}) is the main result of the present work. It
states that $\Psi$ defined above is an exact eigenfunction of the many
electron system in FQHE regime for a   periodically extended Coulomb
interaction. Note that the solution is obtained for any finite particle
systems satisfying the conditions assumed here. Thus, it can be employed to
investigate the thermodynamic limit. \ The evaluation of the energy per
particle associated to the obtained solution will be pursued in paper II.

Let us briefly comment here about how the large number of
translations entering the definition of the periodic potential in
the growing number of particles limit, is however able produce the
closure relation which was central in obtaining the argued result. \

It can be noted that for $q=1,$ the set of translations of the form
$\mathbf{l}^{\prime}$ transforms the $LLL$ level set of functions
satisfying the  assumed periodic conditions,  in itself. That is,
the number of translations coincides with the number of independent
function in this level which satisfy the  boundary conditions,  and
any translation implements a permutation of the elements of a
special complete basis of eigenfunctions $\phi_{\mathbf{q}}$ defined
in Refs. \cite{ferrari,cabo,cabo1}. \ These orbitals are
eigenfunctions of the magnetic translations on a lattice showing
exactly one unit of flux per unit cell. \ The form of this unit cell
can be arbitrary, the only requirement on it, is to being pierced by
one flux quanta. An important property of those translations which
is argued by example in Refs.\cite{ferrari,cabo1}, is that any
magnetic translation in the $LLL$ is equivalent to a \ shift in the
momentum quantum numbers of the functions in the mentioned basis.
Let us consider such a $one$ $flux$ $quantum$ unit cell as formed by
the vectors $\mathbf{a}_{1}/q$ and $\mathbf{a}_{2}$. \ The number of
such cells in the area of the sample we are considering is exactly
$q(N^{2}/q^{2})$, that is, the number of flux quanta passing through
the sample area. \ In this case, the momenta associated to the
complete set of
functions $\phi_{\mathbf{q}}$ in the chosen sample area are\cite{ferrari,cabo,cabo1}%
\begin{align}
\mathbf{q} &  =\frac{q_{1}}{N}\mathbf{s}_{1}+\frac{q_{2}}{\frac{N}{q}%
}\mathbf{s}_{2,}\nonumber\\
s_{1} &  =-\frac{\mathbf{n}\times\mathbf{a}_{2}}{r_{o}^{2}},\text{ \ }%
s_{2}=-\frac{\mathbf{n}\times\mathbf{a}_{2}}{r_{o}^{2}},\nonumber\\
q_{1} &  =-\frac{N}{2},-\frac{N}{2}+1,...,0,..,\frac{N}{2}-1,\nonumber\\
q_{2} &  =-\frac{N}{2q},-\frac{N}{2q}+1,...,0,..,\frac{N}{2q}-1.
\end{align}

Let us consider now the $(\frac{N}{q})^{2}$ translations generated
by the vectors $\ \mathbf{l}^{\prime}$ when $\mathbf{l}^{\prime}$ is
restricted to be a quantum number of the functions
$\Theta_{\mathbf{l}^{\prime}}$ \ Then, the shifts in  the momenta
$\mathbf{q}$ produced on the functions $\phi _{\mathbf{q}}$ after
acting on them with a translation  $\ \mathbf{l}^{\prime}$ is given
by \cite{cabo1}
\begin{align}
\delta\mathbf{q} &  =-\frac{2e}{\hslash c}\mathbf{A}(\mathbf{l}^{\prime})\nonumber\\
&  =-\frac{2e}{\hslash c}\frac{B}{2}\mathbf{n}\times(\mathbf{l}^{\prime})\nonumber\\
&  =\frac{1}{r_{o}^{2}}\mathbf{n}\times\mathbf{l}^{\prime}\nonumber\\
&  =\frac{1}{r_{o}^{2}}\mathbf{n}\times(\frac{r_{1}}{N}\mathbf{a}_{1}%
+\frac{r_{2}}{N}\mathbf{a}_{2})\nonumber\\
&  =(-q\frac{r_{2}}{N}\mathbf{s}_{1}+\frac{r_{1}}{\frac{N}{q}}\mathbf{s}_{2}).
\end{align}

This expression indicates that when $q>1$ there exist a proper subset of the
functions $\phi_{\mathbf{q}}$ that are closed under all the \ magnetic
translations defining the specially periodic Coulomb operator. This conclusion
follows form the fact that the shifts $\delta\mathbf{q}$ are closed under
addition, but they only have $(\frac{N}{q})^{2}$ values within the\ first
Brillouin zone, of the total number $q(\frac{N}{q})^{2}$ of momenta forming
this zone. Therefore, the existence of a  close proper subset formed by the
functions $\Theta_{\mathbf{l}}$ within the $LLL$ subject to the considered
periodic conditions is allowed.

\subsection{Formula for the energy per particle of the exact eigenstates}

\  Let us employ in what follows the periodicity  conditions for
finding a formula for the energy per particle  in terms of the
$\Theta _{\mathbf{l}}$ functions, for the FQHE eigenstates. Since
the state $\Psi$ is a Slater determinant the total energy in the von
Karman periodicity zone can be expressed in the form
\begin{align}
E &  =\langle\text{ }\Psi\text{ }|\text{ }V_{c}|\text{ }\Psi\text{ }%
\rangle\nonumber\\
&  =\int_{A_{c}}\int_{A_{c}}d\mu(\mathbf{x})d\mu(\mathbf{x}^{\prime}%
)V_{c}(\mathbf{x}-\mathbf{x}^{\prime})g(\mathbf{x},\mathbf{x}^{\prime
}),\label{energy}\\
g_{_{Ne}}(\mathbf{x},\mathbf{x}^{\prime}) &  =\rho(\mathbf{x},\mathbf{x}%
)\rho(\mathbf{x}^{\prime},\mathbf{x}^{\prime})-\rho(\mathbf{x},\mathbf{x}%
^{\prime})\rho(\mathbf{x}^{\prime},\mathbf{x}),\nonumber\\
\rho(\mathbf{x},\mathbf{x}^{\prime}) &  =\sum_{\mathbf{l}}\Theta_{\mathbf{l}%
}(\mathbf{x})\Theta_{\mathbf{l}}^{\ast}(\mathbf{x}^{\prime}),
\end{align}
where $\Theta_{\mathbf{l}}(\mathbf{x})$ as a function of the
coordinate vector $\mathbf{x}$ is naturally defined by
\begin{equation}
\Theta_{\mathbf{l}}(\mathbf{x})=\Theta_{\mathbf{l}}(((\mathbf{x}%
)_{1},(\mathbf{x})_{2}))\equiv\Theta_{\mathbf{l}}(z^{\ast})=\Theta
_{\mathbf{l}}((\mathbf{x})_{1}-i(\mathbf{x})_{2}),
\end{equation}
and as usual, in the Landau gauge the integration measure is given
by
\begin{equation}
d\mu(\mathbf{x})=d\text{ }\mathbf{x}\text{ }\exp(-\frac{((\mathbf{x})_{2}%
)^{2}}{r_{o}^{2}}).
\end{equation}
We also assumed here that the functions $\Theta_{\mathbf{l}}(z^{\ast})$ were
already normalized in the  von Karman periodicity zone.

\ \ Let us consider in what follows the following type of \ integrals%
\begin{equation}
I=\int_{A_{c}}d\mu(\mathbf{x})V_{c}(\mathbf{x}-\mathbf{x}^{\prime}%
)\Theta_{\mathbf{l}}(\mathbf{x)}\Theta_{\mathbf{l}^{\prime}}(\mathbf{x),}%
\label{integral1}%
\end{equation}
which are building elements of the expression for the energy
(\ref{energy}). To start, note that the region of integration is
limited to a cell of the lattice $\mathbf{R}_{q},$which was named
above as $A_{q}$. For this purpose. we can rewrite (\ref{inva1}),
after substituting the explicit form of the \ magnetic translation
invariance in the Landau gauge (\ref{magtrans}), as follows
\begin{align}
\Theta_{\mathbf{l}}(z^{\ast}) &  =T(\mathbf{R}_{q},z^{\ast})\Theta
_{\mathbf{l}}(z^{\ast})\nonumber\\
&  =\exp(-i\frac{(\mathbf{R}_{q})_{2}R_{q}^{\ast}}{2r_{o}^{2}}-i\frac
{(\mathbf{R}_{q})_{2}z^{\ast}}{r_{o}^{2}})\Theta_{\mathbf{l}}(z^{\ast
}+\mathbf{R}_{q}).
\end{align}

Employing this relation for the two functions $\Theta_{\mathbf{l}}$ and
$\Theta_{\mathbf{l}^{\prime}}$ in (\ref{integral1}) and taken into account the
periodicity of the potential in the lattice $\mathbf{R}_{q}$, it follows
\begin{align}
I &  =\int_{A_{c}}d\mu(\mathbf{x})V_{c}(\mathbf{x}-\mathbf{x}^{\prime}%
)\Theta_{\mathbf{l}}(\mathbf{x+R}_{q}\mathbf{)}\Theta_{\mathbf{l}^{\prime}%
}(\mathbf{x+R}_{q}\mathbf{)\times}\nonumber\\
&  \exp(-i\frac{(\mathbf{R}_{q})_{2}(R_{q}^{\ast}-R_{q})}{2r_{o}^{2}}%
-i\frac{(\mathbf{R}_{q})_{2}(z^{\ast}-z)}{r_{o}^{2}})\nonumber\\
&  =\int_{A_{c}}d\mu(\mathbf{x+R}_{q})V_{c}(\mathbf{x}-\mathbf{x}^{\prime
})\Theta_{\mathbf{l}}(\mathbf{x+R}_{q}\mathbf{)}\Theta_{\mathbf{l}^{\prime}%
}(\mathbf{x+R}_{q}\mathbf{).}%
\end{align}

This expression explicitly shows that the integral $I$ performed
over an arbitrary \ cell of the lattice $R_{q}$, give the same
result when performed over any of such cells. \ This property\
directly check that total energy can be evaluated as an integer
limited to the first cell of the lattice $R_{q}$ as
the von Karman conditions implies. Then, the total energy has the formula%
\begin{align}
E_{_{Ne}} &  =\langle\text{ }\Psi\text{ }|\text{ }V_{c}|\text{ }\Psi\text{
}\rangle\nonumber \label{x}\\
&  =\int_{A_{c}}\int_{A_{c}}d\mu(\mathbf{x})d\mu(\mathbf{x}^{\prime}%
)V_{c}(\mathbf{x}-\mathbf{x}^{\prime})g_{_{Ne}}(\mathbf{x},\mathbf{x}^{\prime
}).
\end{align}

\ The evaluation of the energy per particles of the finite systems,
their thermodynamic limits and the pair correlation functions will
be considered in paper II, which is now in preparation.

\subsection{Remarks on the thermodynamic limit}

Let us briefly comment  in this subsection on the \ thermodynamic
limit of the determined exact wavefunctions. As known, in the large
magnetic field regime in which the system is projected in $LLL$
level, the only term in the Hamiltonian is the Coulomb interaction.
Here we assumed von Karman periodic conditions in a magnetic field
(See [\onlinecite{halrez}]), \ with spacial periods \
$\mathbf{L}_{1}/q$ and $\mathbf{L}_{2}/q$ . Let us assume as usual ,
that the systems has a homogeneously distributed jellium which
compensates the net charge of all the particles contained in it. In
this situation,  when the $|\mathbf{L}_{1}|/q=$
$|\mathbf{L}_{2}|/q=L/q=aN/q$ \ tends to be large in the
thermodynamic limit, the interaction between any pair of particles
sitting at arbitrary but finite distances from the origin, should \
approximate the usual Coulomb potential,  as $\frac{L}{q}$ becomes
very much greater than the distance between the mentioned two
particles. This property is expected to be valid because when $L/q$
is very much larger than the distance between the particles, their
interaction potential becomes negligible assumed the compensating
fields of the jellium homogeneous charges are considered (or
equivalently, imposing that the inter particle potential is assumed
as short ranged, by substituting the Coulomb one by a Yukawa
screened potential at large distances). \ In other words, the
jellium charge, should reduce the fields of the far away laying
image charges to be dipolar or higher multipole short ranged
contributions, which should produce vanishing forces in the
thermodynamic limit. Therefore, it can be expected the solution
found in last section, when considered in the limit of a large
number of particles, \ should describe the FQHE systems when the
limit \ $L/q\rightarrow\infty$ \ is taken. In this case of large
samples the following formula for the energy per particle arises
from expression (\ref{x}) \ for the total energy of arbitrary finite systems %
\begin{align}
\epsilon_{FQHE} &  =\lim_{N_{e}\rightarrow\infty}\left(  \frac{1}{N_{e}%
}\langle\text{ }\Psi\text{ }|\text{ }V_{c}|\text{ }\Psi\text{ }\rangle\right)
\nonumber\\
&  =\lim_{N_{e}\rightarrow\infty}\left(  \frac{1}{N_{e}}\int_{A_{c}}%
\int_{A_{c}}d\mu(\mathbf{x})d\mu(\mathbf{x}^{\prime})V_{c}(\mathbf{x}%
-\mathbf{x}^{\prime})g_{_{Ne}}(\mathbf{x},\mathbf{x}^{\prime})\right)  .
\end{align}

\section{Conclusions}

Exact eigenstates of a two dimensional electron gas (2DEG) in a
magnetic field are constructed for finite samples of arbitrary
sizes, and periodic extensions of the Coulomb potential after
restricted to the $LLL$ level.  Magnetic von Karman periodic
boundary conditions are imposed in the Landau gauge. Formulae for
the energy per particle and the two particle density matrix of the
realistic FQHE state  in the thermodynamic limit are also determined
by considering the large limit of finite systems. The found states
are associated to filling factors $\nu=1/q$, for odd $q$. The
results suggests the presence of integrability properties \ in FQHE\
problems. The energy per particle and pair correlation functions of
the states for the finite system and the  thermodynamic limit ones,
will be evaluated in a second work  (paper II) which is now in
preparation. In the case that the energies per particle associated
to the wavefunctions result to be lower than the estimated values
for the Laughlin variational states, they could surprisingly furnish
the ground states of the FQHE problem. However, in the opposite
outcome, the determined wavefunctions could be also helpful in
describing experimentally detected  phases shown by FQHE samples at
variable temperature experiences. In any situation,  the obtained
kind of solution constitutes  a valuable primer of \ exactly
solvable model in the context of the  FQHE theory. The solutions
could  also be exact realizations of the early proposed cooperative
rings of exchange states \cite{kivelson}. \ The evaluations of the
energy per particle and pair correlation functions of the identified
eigenstates \ will be presented in a related work (paper II) which
preparation is in progress.

The special  one particle orbitals constructed here,  were\ also employed (See
Appendix C) to construct ansatz wave functions for a class of low energy
states having a similar structure as the Laughlin's ones. That is, showing
zeros of the form $(z_{i}-z_{j})^{q}$ when the coordinates of the particles
coincide. Then, these states exhibit a similar sort of short range behavior as
the Laughlin states. However, since the basic elements in their definition are
the HF one particle orbitals, we expect that  long range correlations
incorporated in the original HF solution can  remain in the proposed ansatz
states, helping in this way to slightly lower the energy per particle.

\begin{acknowledgments}
One of the authors (A.C.) acknowledges support from various
institutions: the Caribbean Network on Quantum Mechanics, Particles
and Fields (Net-35) of the ICTP Office of External Activities (OEA),
the "Proyecto Nacional de Ciencias B\'{a}sicas" (PNCB) of CITMA,
Cuba, and the Perimeter Institute for Theoretical Physics, Waterloo,
Canada. Both authors acknowledge the support received from Fondecyt,
Grants 1060650 and 7060650, and the Catholic University of Chile.
\end{acknowledgments}

\appendix

\section{Basic definitions}

\subsection{General definitions}

In this work, bold symbols represent vectors with components $\mathbf{v=}%
((\mathbf{v)}_{1},(\mathbf{v)}_{2},(\mathbf{v)}_{3})$ and 2-dimensional
vectors are equivalently defined by the two complex numbers $\ \ \ $%
\[
\ v=(\mathbf{v)}_{1}+i(\mathbf{v)}_{2},\ \ \ \ v^{\ast}=(\mathbf{v)}%
_{1}-i(\mathbf{v)}_{2}.
\]
Let us consider the plane containing the 2DEG. One kind of useful periodicity
planar box employed in previous works consists in a parallelogram of sides
$\mathbf{L}_{1}=N$ $\mathbf{a}_{1}$ and $\mathbf{L}_{2}=N$ $\mathbf{a}_{2}$,
where the unit cell vectors $\mathbf{a}_{1}$ and $\mathbf{a}_{2}$ are defined
by
\begin{align}
\mathbf{a}_{1} &  =a\text{ }(1,0,0),\text{ \ }\mathbf{a}_{2}=a\text{ }%
(\frac{1}{2},\frac{\sqrt{3}}{2},0),\\
\mathbf{n} &  =(0,0,1),\;a=\sqrt{\frac{4\pi q}{\sqrt{3}}}r_{o},
\end{align}
where $q$ is an odd integer, $\mathbf{n}$ is the unit vector normal to the
2DEG plane, and $r_{o}$=$\sqrt{\frac{\hbar c}{|e|B}}$ is the magnetic length.
\ The Landau gauge $\mathbf{A}=B(-x_{2},0,0)$ is employed for the vector
potential in accordance with Ref. [\onlinecite{halrez}] . Note that the cell
with sides $\mathbf{a}_{1}$ and $\ \mathbf{a}_{2}$ intercepts exactly $q$ flux
quanta of the magnetic field $\mathbf{B}=B(0,0,1)$. Also,  $N^{2}$ particles
laying inside the above defined  particular planar box correspond to a filling
factor $\nu=1/q$. \ Note that this box will not be the von Karman periodicity
box considered for determining the exact solutions, which will have sizes
given by $\mathbf{L}_{1}/q$ and $\mathbf{L}_{2}/q$. The electric charge $e$ is
assumed to be negative. For convenience, $N$ \ will be assumed to be an even
positive integer number and also an exact multiple of $q$. \ Is helpful to
define a spatial lattice given by the vectors
\begin{equation}
\mathbf{R}=R_{1}\mathbf{a}_{1}+R_{2}\mathbf{a}_{1},\label{lattice}%
\end{equation}
in which $R_{1}$ and $R_{2}$ are arbitrary positive or negative integers.
\ The reciprocal lattice cell vectors associated to this lattice has unit
vectors
\begin{align}
\mathbf{s}_{1} &  =-\frac{\mathbf{n}\times\mathbf{a}_{2}}{q\text{ }r_{o}^{2}%
},\,\,\mathbf{s}_{2}=\frac{\mathbf{n}\times\mathbf{a}_{1}}{q\text{ }r_{o}^{2}%
},\text{ \ \ }\\
\text{\ }\mathbf{s}_{i}.\mathbf{a}_{j} &  =2\pi\delta_{ij},\text{ \ }i,j=1,2.
\end{align}

The free Hamiltonian of the single planar electron in the Landau gauge will
be
\[
H=\frac{1}{2m}(\mathbf{p}-\frac{e}{c}\mathbf{A})^{2},
\]
where the \ generators of the magnetic translation symmetry operations are%

\begin{align*}
\mathbf{G} &  =\mathbf{\nabla-}\frac{ie}{\hslash c}\mathbf{A+}\frac
{ieB}{\hslash c}\mathbf{n}\times\mathbf{x,}\\
\lbrack\mathbf{G,}H] &  =0,\\
\lbrack(\mathbf{G)}_{i}\mathbf{,(G)}_{j}] &  =\frac{ieB}{\hslash c}%
\epsilon^{ijk}(\mathbf{n})_{k}.
\end{align*}
The structure of the free electron wavefunctions of the $LLL$\ in the Landau
gauge is%
\begin{align*}
\varphi(\mathbf{x}) &  =\phi(z^{\ast})\exp(-\frac{y^{2}}{2r_{o}^{2}}),\text{
\ \ \ }y=(\mathbf{x})_{2}\\
z &  =(\mathbf{x)}_{1}\mathbf{+(x)}_{2}\,i\,,\;\;z^{\ast}=(\mathbf{x)}%
_{1}-\mathbf{(x)}_{2}\,i.
\end{align*}
where $\phi(z^{\ast})$ is an analytic function of \ $z^{\ast}.$ In the axial
gauge, the electron states and the vector potential are
\begin{align*}
\varphi_{A}(\mathbf{x}) &  =\phi_{A}(z^{\ast})\exp(-\frac{\mathbf{x}^{2}%
}{4r_{o}^{2}}),\text{ }\\
\mathbf{A}_{A} &  =\frac{1}{2}\text{\ }B\text{ }\mathbf{n}\times\mathbf{x},
\end{align*}
with $\phi_{A}(z^{\ast})$ an analytic function of $z^{\ast}$ different from
$\phi_{A}(z^{\ast})$. The general connection between wavefunctions in both
gauges, and between the analytic forms associated with the $LLL$ are
\begin{align*}
\varphi_{A}(\mathbf{x}) &  =\exp(\frac{ieB}{2\hslash c}(\mathbf{x)}%
_{1}(\mathbf{x)}_{2})\varphi(\mathbf{x}),\\
\phi_{A}(z^{\ast}) &  =\exp(\frac{(z^{\ast})^{2}}{4r_{o}^{2}})\phi(z^{\ast}).
\end{align*}
Then, the Landau gauge versions of the Hamiltonian $H$, and the symmetry
translations generators $\mathbf{G}$, can be obtained from the similarity transformations%

\begin{align*}
H_{A}  &  =\exp(\frac{ieB}{2\hslash c}(\mathbf{x)}_{1}(\mathbf{x)}_{2}%
)H\exp(-\frac{ieB}{2\hslash c}(\mathbf{x)}_{1}(\mathbf{x)}_{2})\\
&  =\frac{1}{2m}(\mathbf{p}-\frac{e}{c}\mathbf{A}_{A})^{2}=-\frac{\hslash^{2}%
}{2m}\mathbf{\Pi}_{A}^{2}\\
\mathbf{G}_{A}  &  =\exp(\frac{ieB}{2\hslash c}(\mathbf{x)}_{1}(\mathbf{x)}%
_{2})\mathbf{G}\exp(-\frac{ieB}{2\hslash c}(\mathbf{x)}_{1}(\mathbf{x)}_{2})\\
&  =\mathbf{\nabla+}\frac{ie}{\hslash c}\mathbf{A}_{A}\mathbf{,}\\
\mathbf{\Pi}_{A}  &  =\mathbf{\nabla-}\frac{ie}{\hslash c}\mathbf{A}%
_{A}\mathbf{.}%
\end{align*}
These quantities have the following properties%

\begin{align*}
\lbrack(\mathbf{G}_{A})_{i},(\mathbf{\Pi}_{A})_{j}]  &  =0\\
\lbrack(\mathbf{G}_{A})_{i},(\mathbf{G}_{A})_{j}]  &  =\frac{i}{r_{o}^{2}%
}\epsilon_{ilj}(\mathbf{n})_{l},\\
\lbrack(\mathbf{\Pi}_{A})_{i},(\mathbf{\Pi}_{A})_{j}]  &  =-\frac{i}{r_{o}%
^{2}}\epsilon_{ilj}(\mathbf{n})_{l}\\
\lbrack H,\mathbf{G}_{A}]  &  =0\\
T_{A}(\mathbf{L},\mathbf{x})  &  =\exp(\frac{ieB}{2\hslash c}(\mathbf{x)}%
_{1}(\mathbf{x)}_{2})T(\mathbf{L},\mathbf{x})\exp(-\frac{ieB}{2\hslash
c}(\mathbf{x)}_{1}(\mathbf{x)}_{2}).
\end{align*}

Helpful quantities are also the lowering and rising operators $\Pi_{\pm}$ and
$G_{\pm}$ defined by and satisfying the following relations
\begin{align*}
\Pi_{\pm}  &  =(\mathbf{\Pi}_{A}\mathbf{)}_{1}\pm(\mathbf{\Pi}_{A}%
\mathbf{)}_{2}i,\;\\
G_{\pm}  &  =(\mathbf{G}_{A}\mathbf{)}_{1}\pm(\mathbf{G}_{A}\mathbf{)}_{2}i,\\
\lbrack H,\Pi_{\pm}]  &  =\pm\frac{\hslash^{2}}{mr_{o}^{2}}\Pi_{\pm
},\;[H,G_{\pm}]=0,\\
\lbrack\Pi_{+},\Pi_{-}]  &  =\frac{2}{r_{o}^{2}},\;[G_{+},G_{-}]=-\frac
{2}{r_{o}^{2}}.
\end{align*}
They have the explicit forms in terms of the complex representation of the coordinates%
\begin{align*}
\Pi_{+}  &  =2\partial_{z^{\ast}}-\frac{z}{2r_{o}^{2}},\;\;\Pi_{+}%
=2\partial_{z}+\frac{z^{\ast}}{2r_{o}^{2}},\\
G_{+}  &  =2\partial_{z^{\ast}}+\frac{z}{2r_{o}^{2}}, \;\;G_{-}=2\partial
_{z}-\frac{z^{\ast}}{2r_{o}^{2}}.
\end{align*}

Now let us consider a property that becomes useful for projecting the Coulomb
interaction in the $LLL$, since it is convenient for  studying the high
magnetic field limit. \ Firstly, let us express the scalar product of a 2D
coordinate vector $\mathbf{x}$ with a vector $\mathbf{Q}$, as%
\begin{align*}
i\mathbf{Q.x}  &  =-2\mathbf{l}.\frac{ie}{\hslash c}\mathbf{A}_{A}%
\mathbf{(x),}\\
\mathbf{l}  &  =r_{o}^{2}\mathbf{n\times Q.}%
\end{align*}
But the gradient and the axial gauge vector potential can be expressed in
terms of the magnetic translation (MT)\ symmetry $\mathbf{G}_{A}$ and the
operator $\mathbf{\Pi}_{A}$ as
\begin{align*}
\nabla &  =\frac{1}{2}(\mathbf{\Pi}_{A}+\mathbf{G}_{A}),\\
\frac{ie}{\hslash c}\mathbf{A}_{A}\mathbf{(x)}  &  =\frac{1}{2}(\mathbf{G}%
_{A}-\mathbf{\Pi}_{A}).
\end{align*}
Then, the exponential $\exp(i\mathbf{Q.x)}$ can be represented as follows,%

\begin{align*}
\exp(i\mathbf{Q.x)}  &  =\exp(-2\mathbf{l}.\frac{ie}{\hslash c}\mathbf{A}%
_{A}\mathbf{(x)})\\
&  =\exp(\mathbf{l.\Pi}_{A})\exp(-\mathbf{l.G}_{A})\\
&  =\exp(\frac{1}{2}(l^{\ast}\Pi_{+}+l\Pi_{-})\exp(-\frac{1}{2}(l^{\ast}%
G_{+}+lG_{-})\\
&  =\exp(-\frac{l\,l^{\ast}}{4r_{o}^{2}})\exp(\frac{1}{2}l^{\ast}\Pi_{+}%
)\exp(\frac{1}{2}l\,\Pi_{-})\times\\
&  \exp(-\frac{1}{2}(l^{\ast}G_{+}+lG_{-})),
\end{align*}
where the well known identity%

\[
\exp(A+B)=\exp(A)\exp(B)\exp(-\frac{1}{2}[A,B]),
\]
valid when $[A,B]$ commutes with $A$ and with $B,$ has\ been used. Applying
the one particle projection operator in the $LLL$: $\ P_{o}=\sum_{n}%
|n\rangle\langle n|$ \ ($\{|n\rangle\}$ being any complete set in the $LLL$),
it follows that%

\begin{align*}
P_{o}^{(i)}\exp(i\mathbf{Q.x}_{i}\mathbf{)}P_{o}^{(i)} &  =\exp(-\frac
{l\,l^{\ast}}{4r_{o}^{2}})\exp(-\mathbf{l.G}_{A})\\
&  =\exp(-\frac{l\,l^{\ast}}{4r_{o}^{2}})\exp(-\frac{1}{2}(l^{\ast}%
G_{+}+lG_{-}))\\
&  =\exp(-\frac{l\,l^{\ast}}{4r_{o}^{2}})\exp(-\frac{1}{2}(l^{\ast}%
(2\partial_{z_{i}^{\ast}}+\frac{z_{i}}{2r_{o}^{2}})+\\
&  +l(2\partial_{z_{i}}-\frac{z_{i}^{\ast}}{2r_{o}^{2}}))).
\end{align*}

\subsection{$LLL$ projection of the Coulomb interaction operator}

Let us next consider the Fourier expansion of the $2D$ function $f(\mathbf{x}%
)$ which is periodic in a parallelepiped with sides $\mathbf{L}_{1}^{g}$ and
$\mathbf{L}_{2}^{g}$%

\begin{align*}
f(\mathbf{x}) &  =\frac{1}{A_{cell}}\sum_{Q}f(\mathbf{Q})\exp(i\mathbf{Q.x}%
),\\
f(\mathbf{Q}) &  =\int d\mathbf{x}\text{ \ }f(\mathbf{x})\exp(-i\mathbf{Q.x}%
),\\
\delta^{P}(\mathbf{x}-\mathbf{x}^{\prime}) &  =\frac{1}{A_{cell}}\sum_{Q}%
\exp(i\mathbf{Q.(x-x}^{\prime})),\\
A_{cell} &  =\mathbf{n}.\mathbf{L}_{1}^{g}\times\mathbf{L}_{2}^{g}.
\end{align*}
It should noted that here in this subsection $A_{cell},\mathbf{L}_{1}^{g}$,
and $\mathbf{L}_{2}^{g}$ are the area and the unit cell vectors of a general
periodic lattice. Employing this expansion, the Coulomb interaction
Hamiltonian \ can be expressed as follows,%

\begin{align*}
V_{C} &  =\sum_{\substack{i<j\\i,j=1,2,...N_{e}}}\sum_{R_{c}}\frac{e^{2}%
}{|\mathbf{x}_{i}-\mathbf{x}_{j}+\mathbf{R}_{c}|}\\
&  =\frac{1}{A_{cell}}\sum_{i<j}\sum_{\mathbf{Q}}\frac{2\pi e^{2}}%
{|\mathbf{Q}|}\exp(i\mathbf{Q.(x}_{i}\mathbf{-x}_{j}^{\prime})).
\end{align*}
where $\mathbf{R}_{c}$ is the lattice generated by the unit cell
vectors $\mathbf{L}_{1}^{g}$ and $\mathbf{L}_{2}^{g}$ and
$\mathbf{Q}$ are Fourier components of the periodic  potential in
the box with area  $A_{cell}$. But applying on
$V_{C}$ the many-particle projection operator in the $LLL$%

\[
\widehat{P_{o}}(\mathbf{x}_{1}\mathbf{,x}_{2}\mathbf{,...,x}_{N_{e}}%
)=\prod_{i=1}^{N_{e}}P_{o}^{(i)},
\]
the projected Coulomb interaction$\ \ \widehat{V}_{C,A}$ can be written in the
axial gauge as%

\begin{align*}
\widehat{V}_{C,A} &  \equiv\widehat{P_{o}}V_{C}\widehat{P_{o}}=\frac
{1}{A_{cell}}\sum_{i<j}\sum_{\mathbf{Q}}\frac{2\pi e^{2}}{|\mathbf{Q}|}%
P_{o}(\mathbf{x}_{i})\exp(i\mathbf{Q.x}_{i})P_{o}(\mathbf{x}_{i}%
)P_{o}(\mathbf{x}_{j})\exp(-i\mathbf{Q.x}_{j}^{\prime})P_{o}(\mathbf{x}%
_{j}),\\
&  =\frac{1}{A_{cell}}\sum_{i<j}\sum_{\mathbf{l}}\frac{2\pi r_{o}^{2}e^{2}%
}{|\mathbf{l}|}\exp(-\frac{l\,l^{\ast}}{2r_{o}^{2}})\exp(-\mathbf{l.G}%
_{A}(\mathbf{x}_{i}))\exp(\mathbf{l.G}_{A}(\mathbf{x}_{j}))\\
&  =\frac{1}{A_{cell}}\sum_{i<j}\sum_{\mathbf{l}}\frac{2\pi r_{o}^{2}e^{2}%
}{|\mathbf{l}|}\exp(-\frac{l\,l^{\ast}}{2r_{o}^{2}})\exp(\frac{1}{2}(l^{\ast
}(2\partial_{z_{i}^{\ast}}+\frac{z_{i}}{2r_{o}^{2}})+l(2\partial_{z_{i}}%
-\frac{z_{i}^{\ast}}{2r_{o}^{2}})))\times\\
&  \exp(-\frac{1}{2}(l^{\ast}(2\partial_{z_{j}^{\ast}}+\frac{z_{j}}{2r_{o}%
^{2}})+l(2\partial_{z_{j}}-\frac{z_{j}^{\ast}}{2r_{o}^{2}}))),\\
\mathbf{l} &  \mathbf{=}r_{o}^{2}\mathbf{n}\times\mathbf{Q}.
\end{align*}
This operator may be written in the representation given by the analytic
factors by considering the following rules expressing how the translation
operators transform when extracting the exponential factors from the
wavefunctions:
\begin{align*}
2\partial_{z_{j}}-\frac{z_{j}^{\ast}}{2r_{o}^{2}} &  \rightarrow-\frac
{z_{j}^{\ast}}{r_{o}^{2}},\\
2\partial_{z_{j}^{\ast}}+\frac{z_{j}}{2r_{o}^{2}} &  \rightarrow
2\partial_{z_{j}^{\ast}}.
\end{align*}
Then, the projected Coulomb interaction in the axial gauge acquires the form
\begin{align*}
\widehat{V}_{C,A}^{Anal} &  =\frac{1}{A_{cell}}\sum_{i<j}\sum_{\mathbf{l}%
}\frac{2\pi r_{o}^{2}e^{2}}{|\mathbf{l}|}\exp(-\frac{l\,l^{\ast}}{2r_{o}^{2}%
})\exp(l^{\ast}\partial_{z_{i}^{\ast}}-l\frac{z_{i}^{\ast}}{2r_{o}^{2}}%
)\times\\
&  \exp(-l^{\ast}\partial_{z_{j}^{\ast}}+l\frac{z_{j}^{\ast}}{2r_{o}^{2}})\\
&  =\frac{1}{A_{cell}}\sum_{i<j}\sum_{\mathbf{l}}\frac{2\pi r_{o}^{2}e^{2}%
}{|\mathbf{l}|}\exp(-\frac{l\,l^{\ast}}{2r_{o}^{2}})T_{A}(-\mathbf{l,}%
z_{i}^{\ast})T_{A}(\mathbf{l,}z_{j}^{\ast})\\
T_{A}(\mathbf{l},z^{\ast}) &  =\exp(-\frac{l\,l^{\ast}}{4r_{o}^{2}}%
-\frac{l\,z^{\ast}}{2r_{o}^{2}})\exp(L^{\ast}\frac{\partial}{\partial z\ast}),
\end{align*}
where $T_{A}(\mathbf{l},z^{\ast})$ is the translation operator in the axial
gauge acting in the space of analytic functions. \ \ After performing the
many-particle similarity transformation which transforms axial gauge operators
into the Landau gauge, the Coulomb interaction reduced to the $LLL$ in this
latter gauge takes the form%

\begin{align*}
\widehat{V}_{C}^{Landau} &  ={\Large (}\prod_{i=1}^{N_{e}}\exp(-\frac
{(z_{i}^{\ast})^{2}}{4r_{o}^{2}}){\Large )}\widehat{V}_{C,A}^{Anal}%
{\Large (}\prod_{i=1}^{N_{e}}\exp(\frac{(z_{i}^{\ast})^{2}}{4r_{o}^{2}%
}){\Large )}\\
&  =\frac{1}{A_{cell}}\sum_{i<j}\sum_{\mathbf{l}}\frac{2\pi r_{o}^{2}e^{2}%
}{|\mathbf{l}|}\exp(-\frac{l\,l^{\ast}}{2r_{o}^{2}})T(-\mathbf{l,}z_{i}^{\ast
})T(\mathbf{l,}z_{j}^{\ast})\\
&  \equiv V,
\end{align*}
where now the Landau gauge translation operators acting on the analytic parts
of the single particle wavefunctions have the form%

\begin{equation}
T(\mathbf{l},z^{\ast})=\exp(-i\frac{(\mathbf{l})_{2}l^{\ast}}{2r_{o}^{2}%
}-i\frac{(\mathbf{l})_{2}z^{\ast}}{r_{o}^{2}})\exp(l^{\ast}\frac{\partial
}{\partial z^{\ast}}).\label{magtrans}%
\end{equation}

\section{von Karman boundary  conditions}

Let us argue below that the functions $\Theta_{\mathbf{l}}(z^{\ast})$ defined
above are invariant under the translations.%
\begin{align*}
T(\frac{\mathbf{L}_{1}}{q},z^{\ast}) &  =T(N\frac{\mathbf{a}_{1}}{q},z^{\ast
}),\\
T(\frac{\mathbf{L}_{2}}{q},z^{\ast}) &
=T(N\frac{\mathbf{a}_{2}}{q},z^{\ast}).
\end{align*}
This property, implies that those  functions satisfy  von Karman periodicity
condition  under translations \ in the periods $\frac{\mathbf{L}_{1}}{q}%
,\frac{\mathbf{L}_{2}}{q}.$This result can be directly derived from the
relation (\ref{relat}). \

Consider first  the shift in $\frac{\mathbf{L}_{1}}{q}.$ Then, let
us fix the parameters in  (\ref{relat}) in the form
\begin{align}
\mathbf{l}^{\prime} &  =\frac{\mathbf{L}_{1}}{q},\nonumber\\
\mathbf{l}-\mathbf{l}^{\prime} &  =\mathbf{l}-\frac{\mathbf{L}_{1}}%
{q},\nonumber\\
\lbrack\mathbf{l}-\mathbf{l}^{\prime}]_{red} &  =\mathbf{l,}\nonumber\\
t_{1} &  =\frac{N}{q},t_{2}=0,
\end{align}
where it has been employed that, since
$\mathbf{L}_{1}/q$=$N\mathbf{a}_{1}/q$ \ is a linear combination
with integer coefficients of the vectors $\mathbf{a}_{1}$ and
$\mathbf{a}_{2}$ thanks to the fact that $N$ was supposed as even
and also as a  multiple of $q$. \ Substituting these values in \
(\ref{relat}) directly \ leads to the first von Karman boundary
condition
\begin{equation}
T(\frac{\mathbf{L}_{1}}{q},z^{\ast})\text{ }\Theta_{\mathbf{l}}(z^{\ast
})=\text{ }\Theta_{\mathbf{l}}(z^{\ast}).
\end{equation}

Similarly for a shift in $\frac{\mathbf{L}_{2}}{q},$ the required parameters
in \ (\ref{relat}) become%

\begin{align}
\mathbf{l}^{\prime} &  =\frac{\mathbf{L}_{2}}{q},\nonumber\\
\mathbf{l}-\mathbf{l}^{\prime} &  =\mathbf{l}-\frac{\mathbf{L}_{2}}%
{q},\nonumber\\
\lbrack\mathbf{l}-\mathbf{l}^{\prime}]_{red} &  =\mathbf{l,}\nonumber\\
t_{1} &  =0,t_{2}=\frac{N}{q},
\end{align}
which \  after performing a little more complicate algebra implies the second
von Karman boundary condition
\begin{equation}
T(\frac{\mathbf{L}_{2}}{q},z^{\ast})\text{ }\Theta_{\mathbf{l}}(z^{\ast
})=\text{ }\Theta_{\mathbf{l}}(z^{\ast}).
\end{equation}

Therefore, the functions $\Theta_{\mathbf{l}}$ are invariant under
magnetic translations in the lattice vectors pertaining to $R_{q}$ as%
\begin{equation}
T(\mathbf{R}_{q},z^{\ast})\Theta_{\mathbf{l}}(z^{\ast})=\Theta_{\mathbf{l}%
}(z^{\ast}).\label{inva1}%
\end{equation}

\subsection{An auxiliary relation}

Let us consider the product of Theta functions%

\[
\Omega(v^{\ast}+m^{\ast})=\prod_{R}\theta_{1}(\frac{\pi}{L}(v^{\ast}-(R^{\ast
}-m^{\ast}))|-\tau^{\ast}),
\]
where $m^{\ast}$ is the complex representation of the vector $\mathbf{m}%
=m_{1}\,\mathbf{a}_{1}\mathbf{+}m_{2}\,\mathbf{a}_{2},$ and the product over
$R$ runs as before over the vectors%

\begin{align}
\mathbf{R}  &  =R_{1}\mathbf{a}_{1}+R_{2}\mathbf{a}_{2},\\
R_{1},R_{2}  &  =-\frac{N}{2},...,-1,0,1,...\frac{N}{2}-1.
\end{align}
The simple properties of the Theta functions under shifts in vectors of the
form $R_{1}\mathbf{a}_{\mathbf{1}},$ and the even character of $N,$ make
$\Omega$ independent of $R_{1}.$ \ Therefore%

\begin{align}
\Omega(v^{\ast}+m^{\ast})  &  =\Omega(v^{\ast}+m_{2}a_{2}^{\ast})\nonumber\\
&  =\prod_{R}\theta_{1}(\frac{\pi}{L}(v^{\ast}-(R_{1}a+(R_{2}-m_{2}%
)a_{2}^{\ast}))|-\tau^{\ast}).
\end{align}
Now, it can be noted that $m_{2}$ can be interpreted as a shift in the
arguments $R$ of the factors, defining a modified set of arguments $R^{\prime
}$. \ Then, \ let us write the product $\mathcal{P}$ of all the factors in
which the modified arguments $R^{\prime}$ are not appearing in the original
set $R$. Assuming $m_{2}>0,$ this product can be given the form%

\begin{align}
\mathcal{P}  &  =\prod_{R_{1}}\prod_{j=1}^{m_{2}}\theta_{1}(\frac{\pi}%
{L}(v^{\ast}-(R_{1}a-\frac{N}{2}a_{2}^{\ast}-ja_{2}^{\ast}))|-\tau^{\ast})\\
&  =\prod_{R_{1}}\prod_{j=1}^{m_{2}}\theta_{1}(\frac{\pi}{L}(v^{\ast}%
-(R_{1}a+(\frac{N}{2}-j)a_{2}^{\ast}))+\pi\tau^{\ast}|-\tau^{\ast}).\nonumber
\end{align}
After iterating the usual formula of the $\theta_{1}$ Theta function
under a single shift in $\pi\tau^{\ast},$ the transformation rule
when the argument is
shifted an integer $k$ times $\pi\tau^{\ast},$ can be written in the form \cite{abram}%

\begin{align}
\theta_{1}(u^{\ast}+k\pi\tau^{\ast}|-\tau^{\ast})  &  =\frac{(-1)^{k}%
}{q^{k^{2}}(\tau^{\ast})}\exp(2\,i\,k\,u^{\ast})\theta_{1}(u^{\ast}%
|-\tau^{\ast}))\theta_{1}(u^{\ast}|-\tau^{\ast})\\
q(\tau^{\ast})  &  =\exp(-i\pi\tau^{\ast}),\ k=-\infty,...,-1,0,1,...\infty
.\nonumber
\end{align}
The above expression allows to write $\mathcal{P}$ \ in the form%
\begin{align*}
\mathcal{P}  &  =\prod_{R_{1}}\prod_{j=1}^{m_{2}}\frac{(-1)}{q(\tau^{\ast}%
)}\exp(\frac{2i\pi}{L}(v^{\ast}-(R_{1}a+(\frac{N}{2}-j)a_{2}^{\ast})))\times\\
&  \theta_{1}(\frac{\pi}{L}(v^{\ast}-(R_{1}a+(\frac{N}{2}-j)a_{2}^{\ast
}))|-\tau^{\ast})\\
&  =\exp(+2i\pi m_{2}\frac{v}{a}^{\ast}-\frac{2i\pi m_{2}}{N}\sum
_{R1=-\frac{N}{2}}^{\frac{N}{2}-1}R_{1}a+\\
&  +2i\pi\tau^{\ast}\sum_{j=1}^{m_{2}}\,j\,)\times\prod_{R_{1}}\prod
_{j=1}^{m_{2}}\theta_{1}(\frac{\pi}{L}(v^{\ast}-(R_{1}a+(\frac{N}{2}%
-j)a_{2}^{\ast}))|-\tau^{\ast}).
\end{align*}
Then, the use of the relations%
\begin{align*}
\sum_{R1=-\frac{N}{2}}^{\frac{N}{2}-1}R_{1}  &  =-\frac{N}{2},\\
\sum_{j=1}^{m_{2}}\,j\,  &  =\frac{m_{2}(m_{2}+1)}{2},
\end{align*}
reduces the expression for $\mathcal{P}$ to the same product in which the
$m_{2}$ shift is absent, after multiplication by an exponential factor, as follows%

\begin{align}
\mathcal{P} &  =\prod_{R_{1}}\prod_{j=1}^{m_{2}}\theta_{1}(\frac{\pi}%
{L}(v^{\ast}-(R_{1}a-\frac{N}{2}a_{2}^{\ast}-ja_{2}^{\ast}))|-\tau^{\ast})\\
&  =\exp(i\pi m_{2}+2i\,\pi\,m_{2}\frac{v}{a}^{\ast}+i\,\pi\,\tau^{\ast}%
m_{2}(m_{2}+1)\,)\times\nonumber\\
&  \prod_{R_{1}}\prod_{j=1}^{m_{2}}\theta_{1}(\frac{\pi}{L}(v^{\ast}%
-(R_{1}a+(\frac{N}{2}-j)a_{2}^{\ast}))|-\tau^{\ast}).\nonumber
\end{align}
Then, the following resulting relation arises%
\begin{equation}
\Omega(v^{\ast}+m^{\ast})=\exp(i\pi m_{2}+2i\,\pi\,m_{2}\frac{v}{a}^{\ast
}+i\,\pi\,\tau^{\ast}m_{2}(m_{2}+1)\,)\,\Omega(v^{\ast}),\label{aux1}%
\end{equation}
checking that shifts in the vectors $m^{\ast}$ of the functions $\Omega$
reproduce the same function times an exponential factor.

\section{ Ansatz states for the FQHE problem showing translation symmetry
breaking at $\nu=\frac{1}{q}$}

Let us employ in this section the composite fermion-like basis defined in
Refs.\cite{alamos1,alamos2,ictp} and in (\ref{composf}), to construct a
special class of many-particle state associated with the fractional filling
factor $1/q.$ These wavefunctions show a two particle pair correlation
function having the known optimized behavior $g(\mathbf{x}_{1}-\mathbf{x}%
_{2})\sim$\ $c$($z_{1}^{\ast}-z_{2}^{\ast})^{q}$ which is helpful in
reducing the correlation energy. However, the original source of the
definitions for the basis functions $\chi_{\mathbf{k}_{i}}$ (which
come from HF single particle solutions after extracting the fixed
position zeros) leads to expecting the presence of long range
correlations further diminishing the energy per particle
\cite{alamos1,alamos2,ictp}. \ \ The ansatz states will be
simply defined by%

\begin{align*}
\Psi_{G}(z_{1}^{\ast},z_{2}^{\ast},...z_{N_{e}}) &  =Det[\Theta_{\mathbf{l}%
_{i}}(z_{j}^{\ast})]^{q},\\
Det[\Theta_{\mathbf{l}_{i}}(z_{j}^{\ast})] &
=\sum_{P}(-1)^{p}\Theta
_{\mathbf{l}_{1^{P}}}(z_{1}^{\ast})\Theta_{\mathbf{l}_{2^{P}}}(z_{2}^{\ast
})...\Theta_{\mathbf{l}_{N_{e}^{P}}}(z_{N_{e}}^{\ast})
\end{align*}
These wavefunctions satisfy the boundary conditions%
\begin{align*}
T(\mathbf{L}_{1}/q,z_{i}^{\ast})\Psi_{G}(z_{1}^{\ast},z_{2}^{\ast}%
,...z_{N_{e}}^{\ast}) &  =\Psi_{G}(z_{1}^{\ast},z_{2}^{\ast},...z_{N_{e}}^{\ast}),\\
T(\mathbf{L}_{2}/q,z_{i}^{\ast})\Psi_{G}(z_{1}^{\ast},z_{2}^{\ast},...z_{N_{e}}^{\ast})
&  =\Psi_{G}(z_{1}^{\ast},z_{2}^{\ast},...z_{N_{e}}^{\ast}), \,
i=1,...,N_{e},
\end{align*}
which directly follow from the boundary conditions for the functions
$\Theta_{\mathbf{l}}$.

\end{document}